\documentclass[conference]{IEEEtran}  

\usepackage{graphicx}
\usepackage{cite}
\usepackage[cmex10]{amsmath}
\usepackage{amssymb}
\usepackage{amsthm}
\usepackage{url}
\usepackage{bm}
\usepackage{color}
\usepackage{subfigure}

\newtheorem{theorem}{Theorem}

\newcommand{\bmrm}[1]{\bm{{\rm #1}}}
\newcommand{\etal}{{\it et al.}~}

\newcommand{\MN}{{\mathrm{MN}}}
\newcommand{\x}{{\bm{x}}}
\newcommand{\y}{{\bm{y}}}
\newcommand{\nS}{\mathsf{S}}
\newcommand{\nR}{\mathsf{R}}
\newcommand{\nD}{\mathsf{D}}
\newcommand{\esr}{\epsilon_{\mbox{\tiny \sf SR}}}
\newcommand{\esd}{\epsilon_{\mbox{\tiny \sf SD}}}
\newcommand{\erd}{\epsilon_{\mbox{\tiny \sf RD}}}
\newcommand{\subS}{\mbox{\tiny \sf S}}
\newcommand{\subR}{\mbox{\tiny \sf R}}
\newcommand{\subD}{\mbox{\tiny \sf D}}
\newcommand{\AEPRsize}{0.46}

\definecolor{mygreen}{rgb}{0.0,0.5,0.0}
\newcommand{\revA}[1]{\textcolor{black}{#1}}    
\newcommand{\revB}[1]{\textcolor{black}{#1}}  
\newcommand{\self}[1]{\textcolor{black}{#1}}   
\begin{document}
%
\title{Spatially Coupled LDPC Codes for Decode-and-Forward in Erasure Relay Channel}
 \author{
 \IEEEauthorblockN{Hironori Uchikawa, Kenta Kasai and Kohichi Sakaniwa}
 \IEEEauthorblockA{Dept.\ of Communications and Integrated
  Systems\\Tokyo Institute of Technology\\152-8550 Tokyo, JAPAN\\
  Email: \{uchikawa, kenta, sakaniwa\}@comm.ss.titech.ac.jp}
  }


\maketitle

\begin{abstract}
We consider spatially-coupled protograph-based LDPC codes for \self{the} three
terminal erasure relay channel.  
It is observed that \self{BP threshold value}, the maximal erasure 
probability of the channel for which
decoding error probability converges to zero, 
of \self{spatially-coupled codes}, \revA{in particular} 
\self{spatially-coupled MacKay-Neal} code, is close to the
theoretical limit for the relay channel.  
\self{Empirical results suggest} that spatially-coupled protograph-based LDPC
codes have great potential to achieve theoretical limit of a general
relay channel.

\end{abstract}


%
\IEEEpeerreviewmaketitle

\section{Introduction}
\label{sec:intro} 
Felstr\"{o}m and Zigangirov constructed the time-varying periodic
\revA{Low-Density Parity-Check (LDPC)}
convolutional codes from LDPC block codes \cite{zigangirov99}.  
Surprisingly, the LDPC convolutional codes outperform the
constituent underlying LDPC block codes.  
Recently, Kudekar \etal rigorously proved such decoding performance
improvement over binary erasure channels (BEC) and showed that the terminated LDPC
convolutional coding increases the belief propagation (BP) threshold up
to the maximum a-priori (MAP) threshold of the underlying block code.
This phenomenon is called {\it threshold saturation} \cite{kudekar2011it}.
A protograph of an LDPC convolutional code can be seen that a spatially
coupled protograph of the underlying LDPC code, hence Kudekar \etal named
this code spatially-coupled protograph-based LDPC code.

Spatially-coupled protograph-based LDPC codes, composed of many
identical protographs coupled with their neighboring protographs, 
have recently attracted much attentions.
\self{The threshold saturation phenomenon is observed 
not only for the BEC, but also for general binary memoryless symmetric
(BMS) channels  \cite{kudekar2010bms}.
It is expected that} the spatially-coupled protograph-based LDPC codes
achieve universally the capacity of the BMS channels under BP decoding. 
Such universality is not possessed by polar codes  \cite{arikan2009it}
or irregular LDPC codes  \cite{richardson01design}. 
Depending on the channel, frozen bits need to be optimized for 
polar codes and degree distributions need to be optimized
for irregular LDPC codes. 
Therefore, it is expected that the spatially-coupled protograph-based
LDPC codes \self{are able to be} applied to many other problems in communications.

Recently, Kudekar and Kasai showed empirical evidences that the BP
threshold value of the spatially-coupled protograph-based LDPC codes is
approaching the theoretical limit for a class of channels with memory
\cite{kudekar2011isit-mem} and the Shannon threshold over multiple
access channels \cite{kudekar2011isit-mac}.  

\revB{
MacKay-Neal (MN) codes \cite{mn_code} are non-systematic two-edge type LDPC
codes \cite{ru04met}. 
The MN codes are conjectured to achieve
the capacity of BMS channels under maximum likelihood decoding. Murayama {\it et
al.}~\cite{PhysRevE.62.1577} and Tanaka {\it et
al.}~\cite{TanakaSaad2003} reported the empirical evidence of the
conjecture for BSC and AWGN channels, respectively by a non-rigorous
statistical mechanics approach known as {\em replica method}.
Recently, Kasai \etal
have shown that spatially-coupled MN codes have the BP
thresholds very close to the Shannon limit of the BEC \cite{kasai11ita}.
}

\revB{It is naturally expected that the same phenomenon
occurs also for transmission over relay channels.
We propose spatially-coupled protograph-based LDPC and MN codes
for DF strategy over erasure relay channels.
BP decoding of joint use of Tanner graphs is presented, and density
evolution analysis gives an empirical evidence that 
spatially-coupled protograph-based MN codes achieve 
theoretical limit of the erasure relay channel.}

The paper is organized as follows. 
\self{Section \ref{sec:erc} introduces the erasure relay channel and the DF
strategy.
Section \ref{sec:scldpc} defines spatially-coupled 
protograph-based LDPC and MN codes. 
Section \ref{sec:sc4relay} describes the density evolution equations. 
The numerical results are presented in Section \ref{005318_5Jun11}.
The} last section will conclude. 


\section{Erasure Relay Channel}
\label{sec:erc}
\subsection{Channel Model}
\label{sec:chan}

\begin{figure}[!t]
\centering
\includegraphics[width=0.4\textwidth]{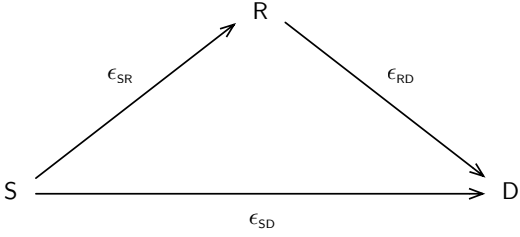}
\caption{Erasure relay channel.}
\label{fig:channel_model}
\end{figure}

\self{We show the erasure relay channel used in this paper in Fig.~\ref{fig:channel_model}. 
The relay channel} comprises of a sender node $\nS$, a destination node $\nD$, and a relay node $\nR$. 
\self{For simplicity,} interferences between the sender and the relay 
transmissions are not considered in this paper, therefore
we can view the above relay channel  as two separate channels.
\self{One is an erasure-broadcast channel from $\nS$ to $\nR$ and $\nD$, and the other is 
a point-to-point erasure channel from $\nR$ to $\nD$.
We denote that the erasure probabilities on the channels from
$\nS$ to $\nR$, from $\nR$ to $\nD$, and from $\nS$ to $\nD$ by $\esr$, $\erd$, 
and $\esd$, respectively.}
This relay channel can be regarded as wireless communication network 
from the viewpoint of higher layer \self{\cite{khalili2004isita}}.

\revA{
\subsection{Capacity of Erasure Relay Channel}
}
\label{sec:cap}
\revA{Denote the coding rate at $\nS$ by $R$. 
\begin{theorem}[Capacity \cite{khalili2004isita}]
The achievable rate region of the erasure relay channel without interferences at $\nD$ is given by: 
\begin{align*}
 R\leq\min\{(1 - \esd \esr), (1 - \esd) + \beta(1-\erd)\},
\end{align*}
where $\beta = 1$ if $R < 1 - \esr$ and $\beta = \esr$ otherwise. 
Since the DF strategy is employed, i.e., $R < 1 - \esr$ in this paper, it holds that $\beta=1.$
The region, therefore, becomes
\begin{align}
 R \leq \min\{(1 - \esr), (1 - \esd) + (1-\erd)\}.
 \label{162148_4Jun11}
\end{align}
\end{theorem}}

\revA{The dashed lines in Fig.~\ref{fig:region.bi.3.6} represents the boundary of the region for fixed coding rate $R=0.5$. 
We will define a code pair used at $\nS$ and $\nR$ in Section
\ref{sec:scldpc}. In Section \ref{sec:sc4relay}, we will investigate the
achievable region of $(\epsilon_{\subS\subR},\epsilon_{\subS\subD})$ for
the code pair. }

\revA{
\subsection{LDPC Coding for DF Strategy}
}
\label{sec:ldpc_df}
\revA{Let $N_{\subS}$ and $N_{\subR}$ are the lengths of codes used at $\nS$ and $\nR$, respectively.
We denote the codewords sent from $\nS$ and $\nR$ by 
$\bm{x}\in\{0,1\}^{N_\nS}$ and $\bm{x}'\in\{0,1\}^{N_\nR}$, respectively.
We denote the received words at $\nR$ and $\nD$ from $\nS$ by  $\bm{y}_\nR\in\{0,1,?\}^{N_\nS}$,  $\bm{y}_\nD\in\{0,1,?\}^{N_\nS}$, respectively.
We denote the received words at $\nD$ from $\nR$  by  $\bm{y}'\in\{0,1,?\}^{N_\nR}$.}

\revA{Design of LDPC codes for relay channels with DF strategy was discussed in several papers
\cite{srid2010isit}, \cite{razaghi2007it}, \cite{nguyen2010isit}, \cite{savin2010se}.
The sender $\nS$ sends a codeword $\bm{x}$ encoded by an LDPC code. 
The relay $\nR$ decodes $\bm{x}$. We assume the decoding error probability is arbitrary small. 
This is realized by capacity approaching codes and due to the DF strategy assumption $R < 1 - \esr$. 
Then $\nR$ generates $\bm{x}'$ from $\bm{x}$ using another LDPC code. 
$\bm{x}'$ is transmitted to $\nD$.
$\nD$ decodes the codeword $\bm{x}$ from $\bm{y}_\nD$ and $\bm{y}'$. 
This decoding process at $\nD$ is performed by joint use of Tanner graphs of the two LDPC codes. }

\self{
\section{Spatially-Coupled Protograph-based Codes}
\label{sec:scldpc}
In this section, we define spatially-coupled protograph-based LDPC and MN codes, respectively. 
}

\self{
\subsection{Protograph-based Codes}
Protograph-based codes are defined by the Tanner graphs lifted from
relatively small graphs called {\it protographs} \cite{thorpe-JPL2003}\cite{div09jsac}.
Protographs are defined by non-negative integer matrices called {\itshape
base-matrices} \cite{lentmaier09ita}.  Let us assume we are given a
base-matrix $\bmrm{B}\in (\mathbb{Z}^{+})^{m_p\times n_p}$.  The
parity-check matrix is obtained by replacing each entry of
$\bmrm{B}(i,j)$ with a $q \times q$ binary matrix which is the sum of
$\bmrm{B}(i,j)$-times randomly chosen $q \times q$ permutation matrices
over $\mathrm{GF}(2)$.  Note that the zero entry of $\bmrm{B}$ is
replaced with a $q\times q$ all-zero binary matrix.  This lifting
process of matrices keeps the weight of columns and rows the same.
}

\self{
\subsection{Spatially-Coupled Protograph-based Codes}
We define a spatially-coupled base-matrix $\bmrm{B}_{[0,L-1]}$ from a given base-matrix $\bmrm{B}\in(\mathbb{Z}^{+})^{m_p\times n_p}$.
Let $L$ be a non-negative integer, which is referred to as {\itshape coupling number}. 
We define $\bmrm{B}_{[0,L-1]}$ as follows. 
\begin{align*}
 \bmrm{B}_{[0,L-1]} &= 
  \begin{bmatrix}
   \bmrm{B}_{0} &  \\
   \vdots & \ddots & \\
   \bmrm{B}_{d} & & \bmrm{B}_{0} \\
    & \ddots & \vdots \\
    &  & \bmrm{B}_{d}
  \end{bmatrix}
\in(\mathbb{Z}^{+})^{m_p(L+d)\times n_pL}, 
\end{align*}
where $\bmrm{B}_{0}, \dotsc, \bmrm{B}_{d}\in(\mathbb{Z}^{+})^{m_p\times n_p}$ are 
non-negative integer matrices chosen so that 
\begin{align*}
 \sum_{i=0}^{d} \bmrm{B}_{i} = \bmrm{B}\in(\mathbb{Z}^{+})^{m_p\times n_p}, 
\end{align*}
for some $d$.
These matrices are referred to as {\it spreading} base-matrices. 
}

\self{
\subsection{Spatially-Coupled Protograph-based ($l,r,L$)-regular LDPC Codes}
We define the base-matrix of protograph-based ($l$,$r$)-regular LDPC codes as 
\begin{align*}
 \bmrm{B}^{(l,r)}&:=[l,\cdots,l]\in (\mathbb{Z}^{+})^{1\times k},
\end{align*}
where we assumed $r=kl$ for some integer $k$, for simplicity. 
Spatially-coupled protograph-based ($l,r,L$)-regular LDPC codes are 
defined as protograph-based codes defined by spreading base-matrices 
\begin{align*}
 \bmrm{B}_i^{(l,r)}=[1,\dotsc,1]\in (\mathbb{Z}^{+})^{1\times k}\text{ for }0\le i\le l-1. 
\end{align*}
}

\self{
The design rate $R^{(l,r,L)}$ of the spatially-coupled protograph-based 
($l,r,L$)-regular LDPC codes is given by
\begin{align}
 R^{(l,r,L)}= 1 - \frac{(L+l-1)}{L k} = R^{(l,r)} -
 \frac{(1-R^{(l,r)})(l-1)}{L}.
 \label{eqn:R_lrL}
\end{align}
$R^{(l,r)}=1-1/k$ is the design rate of the underlying code.
$R^{(l,r,L)}$ converges to $R^{(l,r)}$ as increasing $L$ with gap $O(1/L)$. 
We use bits corresponding to the leftmost column of $\bmrm{B}^{(l,r)}$
and $\bmrm{B}_i^{(l,r)}$ for $i = 0, \dotsc, l-1$
as information bits.
}

\revB{
\subsection{Spatially-Coupled Protograph-based ($l,r,g,L$)-MN Codes}
We define the base-matrix of protograph-based ($l,r,g$)-MN codes as 
\begin{align*}
 \bmrm{B}^{\MN(l,r,g)}&:=
\begin{bmatrix}
 r&1&\cdots&1\\
 \vdots&\vdots&\ddots&\vdots\\
 r&1&\cdots&1
\end{bmatrix}
\in (\mathbb{Z}^{+})^{g\times g+1},
\end{align*}
where we assumed $l=gr$, for simplicity. 
MN codes have punctured nodes, therefore the bits 
corresponding to the leftmost column of
$\bmrm{B}^{\MN(l,r,g)}$ are punctured.}

\revB{
Spatially-coupled protograph-based ($l,r,g,L$)-MN codes are defined as protograph-based
codes defined by spreading base-matrices 
$\bmrm{B}_i^{\MN(l,r,g)}$ for $i = 0, \dotsc, g-1$ as follows
\begin{align*}
 \bmrm{B}_i^{\MN(l,r,g)}&=
\begin{bmatrix}
 \bm{0}_{i-1\times g+1}\\
 \bmrm{b}_i\\
 \bm{0}_{g-i\times g+1}
\end{bmatrix},\\
 \bmrm{b}_i&=[r-1,\bm{0}_{1\times g-i},\bm{1}_i]\text{ for }1\le i\le g-1, \\
 \bmrm{B}_0^{\MN(l,r,g)}&=\bmrm{B}^{\MN(l,r,g)}-\sum_{i=1}^{g-1}\bmrm{B}_i^{\MN(l,r,g)},
\end{align*}
where $\bm{0}_{a\times b}$ represents an $a\times b$ all-zero matrix and
$\bm{1}_{i}$ represents an all-one row vector of length $i$.
}

\revB{The design rate $R^{\MN(l,r,g,L)}$ of the spatially-coupled
protograph-based ($l,r,g,L$)-MN codes is given by
\begin{align}
 R^{\MN(l,r,g,L)}&= 1 - \frac{(g+1)L-(gL+g-1)}{L(g+1)-L} \nonumber \\
&= R^{\MN(l,r,g)} - \frac{1-R^{\MN(l,r,g)}}{L}.
\label{eqn:R_lrgL}
\end{align}
$R^{\MN(l,r,g)}=1/g$ is the design rate of the underlying code.
$R^{\MN(l,r,g,L)}$ converges to $R^{\MN(l,r,g)}$ as increasing $L$ with gap $O(1/L)$. 
}

\revB{We use bits corresponding to the leftmost column of
$\bmrm{B}^{\MN(l,r,g)}$ and $\bmrm{B}_i^{\MN(l,r,g)}$
for $i = 0, \dotsc, g-1$ as information bits. }

\revB{
\subsection{Relay Channel Coding via Spatially-Coupled Protograph-based Codes}
As explained in Section \ref{sec:ldpc_df}, we use two LDPC codes for coding at $\nS$ and $\nR$. 
We propose a relay channel coding scheme via spatially-coupled
protograph-based codes in the following way. 
}

\revB{
The sender $\nS$ encodes the information bits into $\bm{x}$ with a 
spatially-coupled ($l,r,L$)-regular LDPC (resp.~($l,r,g,L$)-MN) code 
defined by a base-matrix $\bmrm{B}^{(l,r)}_{[0,L-1]}$ (resp.~$\bmrm{B}^{\MN(l,r,g)}_{[0,L-1]}$). 
The relay $\nR$ decodes $\bm{x}$ from $\bm{y}_{\nR}$ and encodes the
information bits into $\x'$ with another 
spatially-coupled ($l,r,L$)-regular LDPC (resp.~($l,r,g,L$)-MN) code 
defined by the same base-matrix $\bmrm{B}^{(l,r)}_{[0,L-1]}$ (resp.~$\bmrm{B}^{\MN(l,r,g)}_{[0,L-1]}$). 
}

\begin{figure}[!t]
\centering
\includegraphics[width=0.45\textwidth]{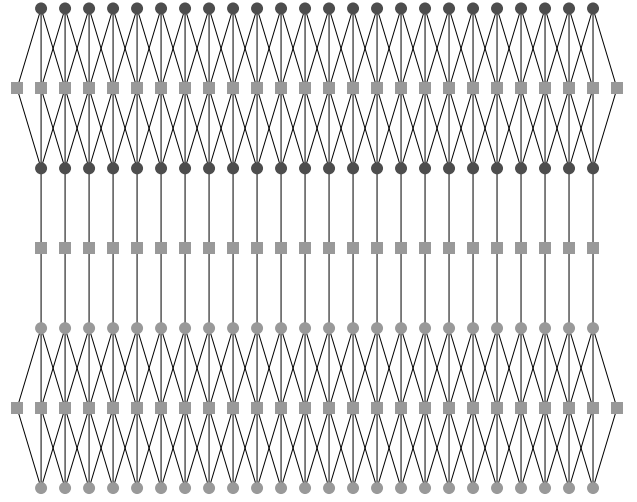}
\caption{
 \self{Protograph of spatially-coupled protograph-based (3,6,24)-regular LDPC codes.}
 The lower protograph with gray circle nodes corresponds to the \self{code
 used at} $\nS$, and the upper protograph
 with dark circle nodes corresponds to the \self{code used at} $\nR$. 
 The lower and upper protographs
 will be connected at $\nD$ \self{for jointly BP decoding}. 
 The channel parameters of \self{the gray circle} nodes
 are $\esd$ and those of \self{the dark circle} nodes are $\erd$.
}
\label{fig:screg}
\end{figure}

\begin{figure}[!t]
\centering
\includegraphics[width=0.48\textwidth]{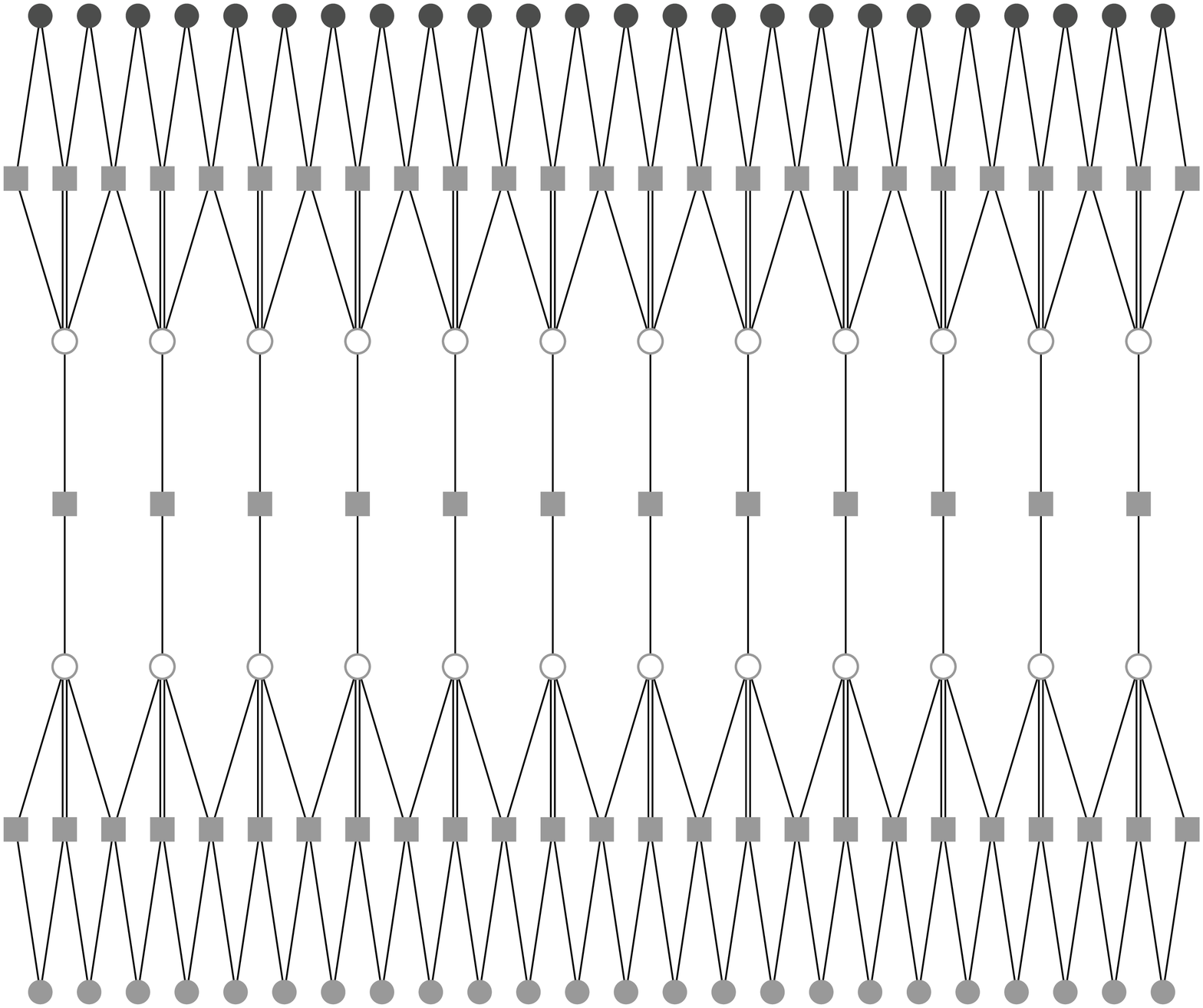}
\caption{
\self{Protograph of spatially-coupled protograph-based (4,2,2,12)-MN codes.}
The lower protograph with gray circle nodes corresponds to the \self{code used
 at} $\nS$, and the upper protograph
with dark circle nodes corresponds to the \self{code used at $\nR$}.
 The lower and upper protographs
 will be connected at $\nD$ \self{for jointly BP decoding}. 
 The channel parameters of \self{the gray circle} nodes
 are $\esd$ and those of \self{the dark circle} nodes are $\erd$.
}
\label{fig:blscmn}
\end{figure}

\revB{
The destination $\nD$ decodes $\x$ from $\y_{\nD}$ and $\y'$ by
BP decoding. The BP decoding algorithm is performed on a Tanner graph which represents the
two codes. The joint Tanner graph is obtained by connecting information
variable nodes in the two codes with check nodes of degree 2.  
For example, the joint protograph of spatially-coupled 
(3,6,24)-regular LDPC and (4,2,2,12)-MN codes are depicted
in Figs.~\ref{fig:screg} and ~\ref{fig:blscmn}, respectively.
}

\revA{
\section{Density Evolution Analysis}
\label{sec:sc4relay} 
The BP decoder \cite{RU08mct} iteratively exchanges
messages $\in \{0,1,?\}$ between variable nodes and check nodes in the
Tanner graphs.  For transmissions over the BEC, the density evolution
allows us to predict the message erasure probability at each iteration
round.
}

\revA{
Let us assume we are given two protograph-based codes defined by
a spatially-coupled base-matrix $\bmrm{B}$. We refer to the edges in the
Tanner graph corresponding to the base-matrix entry $\bmrm{B}(i,j)$ as
edges at {\it section} $(i,j)$.  Let $y^{(\ell)}_{i,j}$ denote the probability
that the messages from check nodes to variable nodes along the edges at
section $(i,j)$ are ``$?$'' at iteration $\ell$.  Similarly, we define
$x^{(\ell)}_{i,j}$ as the probability that the messages from variable
nodes to check nodes along the edges at section $(i,j)$ are ``$?$'' at
iteration $\ell$.  The messages at the 0-th round are initialized with
channel outputs.  It follows that $x^{(0)}_{i,j}=\epsilon_j$, where
$\epsilon_j$ is defined by $\epsilon_j=\esd$ (reps.~$\erd$) if the bits
are transmitted by $\nS$ (resp.~$\nR$) and corresponding to the $j$-th
column of the base-matrix entry $\bmrm{B}$ are not punctured, and
$\epsilon_j=1$ otherwise.
}

\revA{
A message sent from a check node is ``$?$'' if and only if at least one
of the incoming messages are ``$?$''.  Consequently, we have
\begin{align*}
 y^{(\ell)}_{i,j} = 1 - (1-x^{(\ell-1)}_{i,j})^{(\bmrm{B}(i,j)-1)} \prod_{j' \neq j}(1-x^{(\ell-1)}_{i,j'})^{\bmrm{B}(i,j')}, 
\end{align*}
for $(i,j)$ such that $\bmrm{B}(i,j)\neq 0$.
}

\revA{
A message sent from a variable node is ``$?$'' if all the incoming
messages and the message from the channel are ``$?$''.  Consequently, we
have
\begin{align*}
 x^{(\ell)}_{i,j} = \epsilon_{j} (y^{(\ell)}_{i,j})^{(\bmrm{B}(i,j)-1)} 
 \prod_{i' \neq i} (y^{(\ell)}_{i',j})^{\bmrm{B}(i',j)},
\end{align*}
for $(i,j)$ such that $\bmrm{B}(i,j)\neq 0$.
}

\revA{
The channel erasure probability pair ($\erd$, $\esd$) is said to be {\it achievable} by the protograph-based codes 
 if $\lim_{\ell\to\infty}x^{(\ell)}_{i,j}=0$ for all $i,j$ such that $\bmrm{B}(i,j)\neq 0$.
}

\section{Numerical Results}
\label{005318_5Jun11}
In this section, we evaluate the achievable ($\erd$, $\esd$) region,
referred to as {\it achievable erasure probability region}, for 
LDPC codes and spatially-coupled protograph-based LDPC and MN codes. 
\revB{We choose $L=128$ so that the difference between $R^{\MN(l,r,g,L)}$
and 0.5 is less than 0.01.}

\subsection{Regular LDPC Codes}
\revA{
The joint base matrix of the joint (3,6)-regular LDPC code 
at $\nD$ is given as 
\begin{align}
 \bmrm{B} = 
   \begin{bmatrix}
    3 & 3 & 0 & 0 \\
    0 & 0 & 3 & 3 \\
    1 & 0 & 1 & 0 \\
   \end{bmatrix}.
 \label{eqn:B1}
\end{align}
The 2 leftmost columns and the other columns correspond $\bm{x}$ and
$\bm{x}'$, respectively. Therefore 
$\epsilon_{j} =\esd$ for $j=0 ~\text{and}~ 1$, and
$\epsilon_{j'} =\esd$ for $j'=2 ~\text{and}~ 3$. 
Note that the indices of matrices start from 0. 
We compute achievable 
erasure probability region using density evolution with $\bmrm{B}$
as shown in Eq. (\ref{eqn:B1}).}

\begin{figure}[!t]
\begin{picture}(200,200)(0,0)
\put(0,0)
{
\put(5,5){\includegraphics[width=\AEPRsize\textwidth]{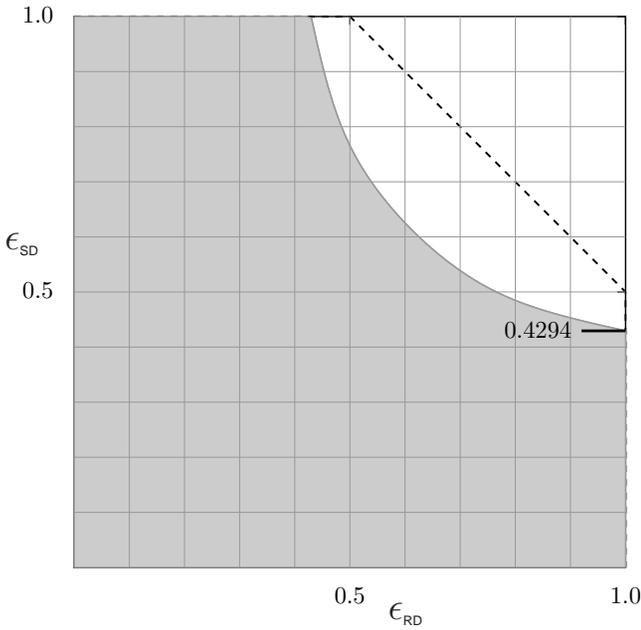}}
\put(0, 140){\rotatebox{0}{\Large $\esd$}}
\put(145, 0){\rotatebox{0}{\Large $\erd$}}
}
\end{picture}
\caption{
Achievable erasure probability region of (3,6)-regular LDPC codes at $\nS$ and $\nR$.
The black dashed line represents the theoretical limit \self{Eq.~\eqref{162148_4Jun11} for rate 0.5.
It can be seen that there is a large gap in the slope region for $0.5 < \erd < 1$, $0.5 < \esd < 1$.}
}
\label{fig:region.bi.3.6}
\end{figure}

Figure \ref{fig:region.bi.3.6} shows the achievable
erasure probability region of (3,6)-regular LDPC codes of rate 0.5 at $\nS$ and $\nR$.
The vertical axis represents $\esd$ and the horizontal axis represents $\erd$.
The black dashed line represents the theoretical limit \self{Eq.~\eqref{162148_4Jun11} for rate 0.5}.
It can be seen that there is a large gap in the slope region for $0.5 < \erd < 1$, $0.5 < \esd < 1$.

\self{
When $\erd = 1$, $\bm{y}'$ is all erased, $\nD$ needs to decode $\bm{x}$ only from $\bm{y}_\nD$. 
Hence, the achievable $\esd$ when $\erd=1$ is equal to the BP
threshold of the (3,6)-regular LDPC code 0.4294.}

\subsection{Split-extended LDPC codes \cite{savin2010se}}
Recently a {\it split-extension} technique for LDPC coding over the
Gaussian relay channels has been developed by Savin \cite{savin2010se}.
The performance over the BEC has not been known, therefore
we evaluate by using achievable erasure probability region
\self{for comparison purpose}.

\self{
Split-extension technique splits the check node of the
protograph to two or more check nodes with variable node 
of degree 2 in order to generate extra parity bits $\x'$
sent from the $\nR$ toward the $\nD$. 
When generated variable nodes of degree 2 are punctured, 
splitted protograph is identical to the original protograph.
Hence $\nR$ sends bits corresponding to the variable nodes 
of degree 2 to the $\nD$. The extra bits can help to 
decode the $\x$ at $\nD$.
}

\revA{
The base matrix of the accumulate repeat jagged accumulate
(ARJA) codes \cite{div-globecom2005} with the split-extension is given as 
\begin{align}
 \bmrm{B} = 
 \begin{bmatrix}
  0 & 0 & 0 & 1 & 0 & 0 & 1 & 0 & 0 \\
  0 & 1 & 0 & 0 & 1 & 1 & 1 & 0 & 0 \\
  1 & 1 & 0 & 1 & 0 & 1 & 0 & 0 & 0 \\
  1 & 1 & 0 & 0 & 0 & 0 & 0 & 1 & 0 \\
  1 & 0 & 0 & 1 & 0 & 0 & 0 & 1 & 1 \\
  1 & 0 & 0 & 0 & 1 & 0 & 0 & 0 & 1 \\
  2 & 0 & 1 & 0 & 0 & 0 & 0 & 0 & 0 \\
  \end{bmatrix}.
 \label{eqn:B2}
\end{align}
The $j$-th columns $j=1,\dotsc,4$ and the $j'$-th columns
$j'=5,\dotsc,8$ and correspond $\bm{x}$ and
$\bm{x}'$, respectively. The leftmost column corresponds
to the punctured bits of ARJA codes.
Therefore 
$\epsilon_{j}=\esd$ for $j=1,2,3,4$, $\epsilon_{j'}=\erd$ for
$j'=5,6,7,8$, and $\epsilon_{0}=1$. 
Note that the indices of matrices start from 0. 
We compute achievable 
erasure probability region using density evolution with $\bmrm{B}$ 
as shown in Eq. (\ref{eqn:B2}).
}

\begin{figure}[!t]
\begin{picture}(200,200)(0,0)
\put(0,0)
{
\put(5,5){\includegraphics[width=\AEPRsize\textwidth]{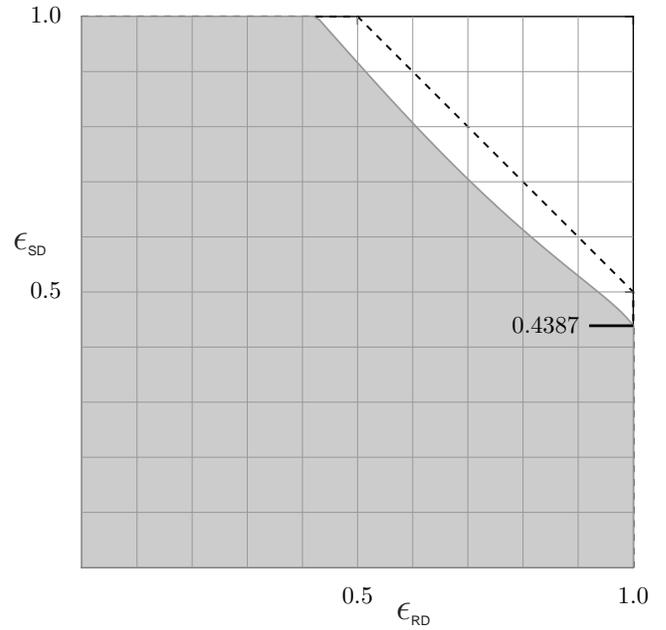}}
\put(0, 140){\rotatebox{0}{\Large $\esd$}}
\put(145, 0){\rotatebox{0}{\Large $\erd$}}
}
\end{picture}
\caption{
Achievable erasure probability region of accumulate repeat jagged accumulate
(ARJA) code \cite{div-globecom2005} with {\it split-extension} technique \cite{savin2010se}.
The black dashed line represents the theoretical limit \self{Eq.~\eqref{162148_4Jun11} for rate 0.5.
It can be seen that there still remains a large gap in the slope region for $0.5 < \erd < 1$, $0.5 < \esd < 1$}.
}
\label{fig:region.se.ARJA}
\end{figure}

Figure \self{\ref{fig:region.se.ARJA}} shows achievable
erasure probability region of ARJA codes with the split-extension.
\self{
When $\erd = 1$, $\bm{y}'$ is all erased, $\nD$ needs to decode $\bm{x}$
only from $\bm{y}_\nD$.
}
Hence, the achievable $\esd$ when $\erd=1$ is equal to the BP
threshold of the ARJA code 0.4387.
It can be seen that there \revA{remains a gap} in the slope region for $0.5 < \erd < 1$, $0.5 < \esd < 1$.

\subsection{Spatially-Coupled Protograph-based $(l,r,L)$-regular LDPC Codes}
\revA{
The joint base matrix of the spatially-coupled (3,6,128)-regular 
LDPC codes at $\nD$ is given as 
\begin{align}
 \bmrm{B} &= 
   \begin{bmatrix}
    \bmrm{B}^{(3,6)}_{[0,127]} & \bmrm{B}^{(3,6)}_{[0,127]} \\
    \bmrm{I'_{128\times256}} & \bmrm{I'_{128\times256}} \\
   \end{bmatrix}  \nonumber \\
 &= 
   \begin{bmatrix}
    1 & 1 & 0 & 0 & \cdots & 1 & 1 & 0 & 0 & \cdots \\
    1 & 1 & 1 & 1 & \cdots & 1 & 1 & 1 & 1 & \cdots \\
    1 & 1 & 1 & 1 & \cdots & 1 & 1 & 1 & 1 & \cdots \\
    0 & 0 & 1 & 1 & \cdots & 0 & 0 & 1 & 1 & \cdots \\
    \vdots & \vdots & \vdots & \vdots & \ddots & \vdots & \vdots & \vdots & \vdots & \ddots \\
    1 & 0 & 0 & 0 & \cdots & 1 & 0 & 0 & 0 & \cdots \\
    0 & 0 & 1 & 0 & \cdots & 0 & 0 & 1 & 0 & \cdots \\
    \vdots & \vdots & \vdots & \vdots & \ddots & \vdots & \vdots &
    \vdots & \vdots & \ddots \\
   \end{bmatrix},
 \label{eqn:B3}
\end{align}
where $\bmrm{I'_{128\times256}}$ is $128 \times 256$ matrix whose $(j,2j)$-th entry is 1 and the other entries are 0. 
These $2j$-th columns correspond to information bits. 
Note that the indices of matrices start from 0. 
The left 256 columns and the other correspond to $\bm{x}$ and
$\bm{x}'$, respectively. 
Therefore $\epsilon_{j}=\esd$ for $j=0,\dotsc,255$,  
and $\epsilon_{j'}=\erd$ for $j'=256,\dotsc,511$. 
We compute achievable 
erasure probability region using density evolution with $\bmrm{B}$
as shown in Eq. (\ref{eqn:B3}).
}

\begin{figure}[!t]
\begin{picture}(200,200)(0,0)
\put(0,0)
{
\put(5,5){\includegraphics[width=\AEPRsize\textwidth]{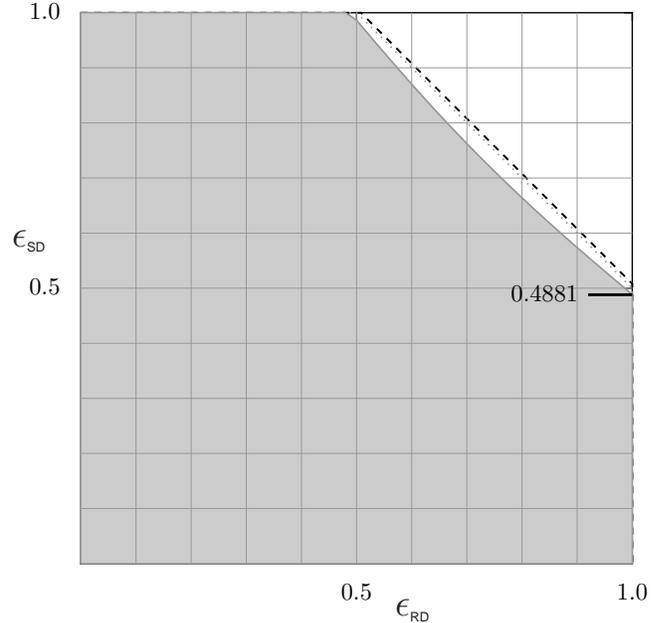}}
\put(0, 140){\rotatebox{0}{\Large $\esd$}}
\put(145, 0){\rotatebox{0}{\Large $\erd$}}
}
\end{picture}
\caption{
Achievable erasure probability region of \self{spatially-coupled
protograph-based (3,6,128)-regular LDPC codes at $\nS$ and $\nR$.}
The black dashed line represents the theoretical limit \self{
 Eq.~\eqref{162148_4Jun11} for design rate 0.4921875} and 
the gray dotted line represents the theoretical limit \self{Eq.~\eqref{162148_4Jun11} for rate 0.5.
There still remains a small gap in the slope region for $0.5 <\erd < 1$,
 $0.5 < \esd < 1$.}
}
\label{fig:region.3.6.128}
\end{figure}

\self{
Figure \ref{fig:region.3.6.128} shows achievable erasure 
probability region of spatially-coupled protograph-based (3,6,128)-regular LDPC
codes at $\nS$ and $\nR$.
The design rate $R^{(3,6,128)}$ is 0.4921875. As $L$ goes to infinity,
$R^{(3,6,L)}$ converges to 0.5 as shown in Eq. (\ref{eqn:R_lrL}).}
The black dashed line represents the theoretical limit for rate 0.4921875 and 
the gray dotted line represents the theoretical limit for rate 0.5.
At the corner point $\erd = 1$ ($\esd = 1$), $\esd$ ($\erd$) is almost
equal to the MAP threshold of (3,6)-regular LDPC code 0.48815.
However, there is still a small gap in the slope region for $0.5 <\erd < 1$, $0.5 < \esd < 1$.


\begin{figure}[!t]
\begin{picture}(200,200)(0,0)
\put(0,0)
{
\put(5,5){\includegraphics[width=\AEPRsize\textwidth]{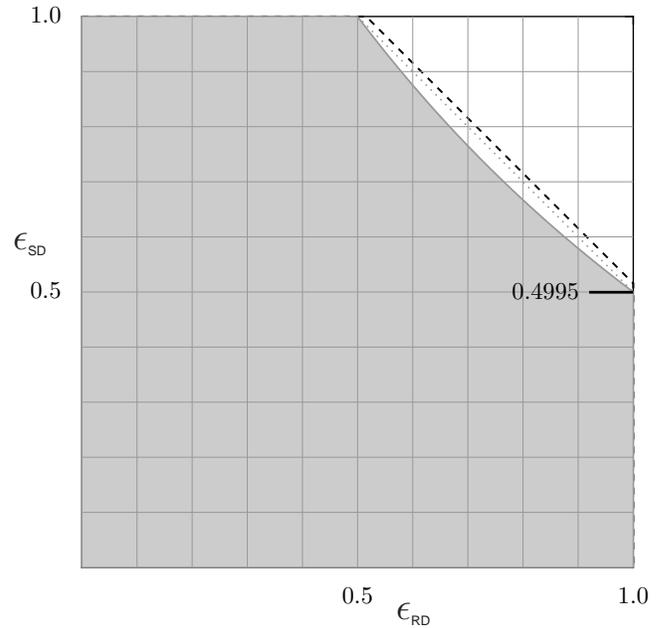}}
\put(0, 140){\rotatebox{0}{\Large $\esd$}}
\put(145, 0){\rotatebox{0}{\Large $\erd$}}
}
\end{picture}
\caption{
Achievable erasure probability region of \self{spatially-coupled (5,10,128)-regular} LDPC
codes at $\nS$ and $\nR$.
The black dashed line represents the theoretical limit \self{Eq.~\eqref{162148_4Jun11} 
for design rate 0.484375} and 
the gray dotted line represents the theoretical limit
 \self{Eq.~\eqref{162148_4Jun11} for rate 0.5.
There still remains a small gap in the slope region for $0.5 <\erd < 1$, $0.5 < \esd < 1$.}
}\label{fig:region.5.10.128}
\end{figure}

\self{
We omit the joint base matrix of 
the spatially-coupled (5,10,128)-regular 
LDPC codes at $\nD$, because it is naturally derived from $\bmrm{B}$
as shown in Eq. (\ref{eqn:B3}).
Figure \ref{fig:region.5.10.128} shows achievable erasure 
probability region of spatially coupled (5,10,128)-regular LDPC
codes at $\nS$ ~and $\nR$.
The design rate $R^{(5,10,128)}$ is 0.484375. 
As $L$ goes to infinity, $R^{(5,10,L)}$ converges to 0.5 as shown in Eq. (\ref{eqn:R_lrL}).
The black dashed line represents the theoretical limit for rate 0.484375 and 
the gray dotted line represents the theoretical limit for rate 0.5.
At the corner point $\erd = 1$ ($\esd = 1$), $\esd$ ($\erd$) is almost
equal to the MAP threshold of (5,10)-regular LDPC code 0.4995.}
However, there \self{still remains a small gap} in the slope region for
$0.5 <\erd < 1$, $0.5 < \esd < 1$.

\self{
$\x'$ from $\nR$ includes repetition bits of
the $\x$ from $\nS$, since the lower part of the joint base matrix 
has rows of weight 2, i.e., [$\bmrm{I'_{128\times256}} \bmrm{I'_{128\times256}}$].
This repetition causes rate loss in the slope region, hence we don't
believe that spatially-coupled ($l,r,L$)-regular LDPC codes can
achieve the theoretical limit of erasure relay channel.
}

\clearpage
\subsection{Spatially-Coupled Protograph-based ($l,r,g,L$)-MN Codes}
The joint base matrix of the spatially-coupled (4,2,2,128)-MN 
codes at $\nD$ is given as 
\begin{align}
 \bmrm{B} &= 
   \begin{bmatrix}
    \bmrm{B}^{\MN(4,2,2)}_{[0,127]} & \bmrm{B}^{\MN(4,2,2)}_{[0,127]} \\
    \bmrm{I''_{128\times384}} & \bmrm{I''_{128\times384}} \\
   \end{bmatrix} \nonumber \\
 &= 
 \begin{bmatrix}
  1 & 1 & 0 & 0 & \cdots & 1 & 1 & 0 & 0 & \cdots \\
  2 & 1 & 1 & 1 & \cdots & 2 & 1 & 1 & 1 & \cdots \\
  1 & 0 & 1 & 1 & \cdots & 1 & 0 & 1 & 1 & \cdots \\
  0 & 0 & 0 & 2 & \cdots & 0 & 0 & 0 & 2 & \cdots \\
  \vdots & \vdots & \vdots & \vdots & \ddots & \vdots & \vdots & \vdots & \vdots & \ddots \\
  1 & 0 & 0 & 0 & \cdots & 1 & 0 & 0 & 0 & \cdots \\
  0 & 0 & 0 & 1 & \cdots & 0 & 0 & 0 & 1 & \cdots \\
  \vdots & \vdots & \vdots & \vdots & \ddots & \vdots & \vdots &
  \vdots & \vdots & \ddots \\
 \end{bmatrix},
 \label{eqn:B4}
\end{align}
where $\bmrm{I''_{128\times384}}$ is $128 \times 384$ matrix 
whose $(j,3j)$-th entry is 1 and the other entries are 0. 
These $3j$-th columns correspond to information bits. 
Note that the indices of matrices start from 0. 
The left 384 columns and the other columns correspond $\bm{x}$ and
$\bm{x}'$, respectively. 
Therefore $\epsilon_j = \esd$ if $j=3t+1, 3t+2$ for $t=0,\dotsc,127$, 
$\epsilon_j' = \erd$ if $j'=3t+1, 3t+2$ for $t=128,\dotsc,255$, and
$\epsilon_{j''} =1$ if $j''=3t$ for $t=0,\dotsc,255$.
We compute achievable 
erasure probability region using density evolution with $\bmrm{B}$
as shown in Eq. (\ref{eqn:B4}).

\begin{figure}[!t]
\begin{picture}(200,200)(0,0)
\put(0,0)
{
\put(5,5){\includegraphics[width=\AEPRsize\textwidth]{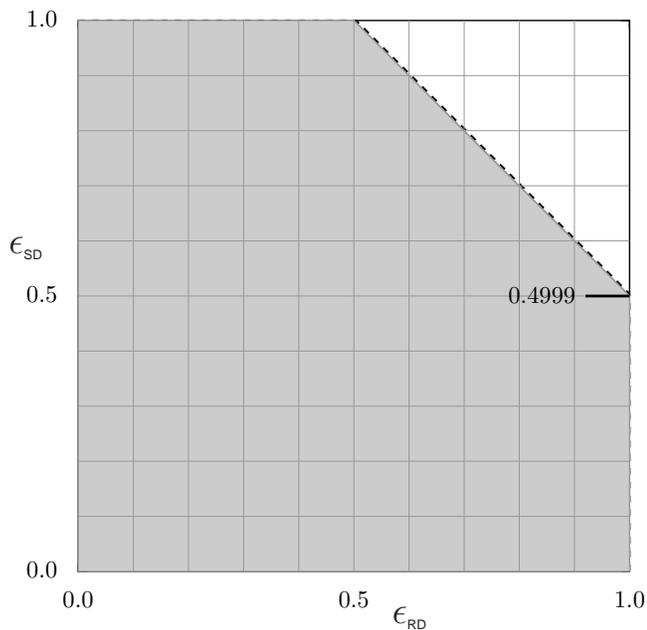}}
\put(0, 140){\rotatebox{0}{\Large $\esd$}}
\put(145, 0){\rotatebox{0}{\Large $\erd$}}
}
\end{picture}
\caption{
Achievable erasure probability region of \self{spatially-coupled
protograph-based (4,2,2,128)-MN} codes at $\nS$ and $\nR$.
The black dashed line represents the theoretical limit
 \self{Eq.~\eqref{162148_4Jun11} for design rate 0.49609375} and 
the gray dotted line represents the theoretical limit \self{Eq.~\eqref{162148_4Jun11} for rate 0.5}.
The achievable region is very close to the theoretical limit. 
However there still remains a very small gap less than $10^{-4}$ between the boundary of the region 
and the theoretical limit. This gap is due to wiggles \cite{kudekar2011it}.
}
\label{fig:region.mn}
\end{figure}

\self{
Figure \ref{fig:region.mn} shows achievable erasure probability region 
of spatially-coupled (4,2,2,128)-MN codes at $\nS$ and $\nR$.
The design rate $R^{\MN(4,2,2,128)}$ is 0.49609375. As $L$ goes to
infinity, $R^{\MN(4,2,2,L)}$ converges to 0.5 as shown in Eq. (\ref{eqn:R_lrgL}).
The black dashed line represents the theoretical limit for rate 0.49609375 and 
the gray dotted line represents the theoretical limit for rate 0.5.
At the corner point $\erd = 1$ ($\esd = 1$), $\esd = 0.4999$ ($\erd=0.4999$) is almost
equal to the point-to-point Shannon limit of rate one half codes.
The boundary of the achievable region is very close to the theoretical limit. 
However there is a very small gap less than $10^{-4}$ between the boundary of the region 
and the theoretical limit. This gap is due to wiggles \cite{kudekar2011it}.
}
%


\section{Conclusion}
\label{sec:conclusion}
\self{
We have designed spatially-coupled protograph-based LDPC and MN codes for erasure relay channels.
It is observed that spatially-coupled protograph-based MN codes approach the theoretical limit. 
We expect that spatially-coupled protograph-based MN codes approach the capacity over the relay
channels} also with other channel impairments, 
such that the binary symmetric relay channels and the Gaussian relay channels.

In the future work, we \self{propose spatially-coupled protograph-based 
MN codes} for Gaussian relay channels and prove
the rate achievability of the spatially\self{-}coupled
protographs-based \self{MN} codes to the theoretical \self{limit}.

\bibliographystyle{IEEEtran}
\bibliography{gotz-ref}

\end{document}